\documentclass[12pt]{article}
\usepackage{epsfig}

\def\beq{\begin{equation}}
\def\eeq#1{\label{#1}\end{equation}}
\def\eeqn{\end{equation}}

\def\beqa{\begin{eqnarray}}
\def\eeqa#1{\label{#1}\end{eqnarray}}
\def\eeqan{\end{eqnarray}}

\let\bar=\overbar

\def\Dslash{\not{\hbox{\kern-4pt $D$}}}
\def\dslash{\not{\hbox{\kern-2pt $\del$}}}

\def\msb{{\bar{\ssstyle M \kern -1pt S}}}

\def\Title#1{\begin{center} {\Large {\bf #1} } \end{center}}

\begin{document}

\Title{Studies of quarkonium at Belle and Belle II}

\bigskip\bigskip

\begin{raggedright}  

{\it Bryan Fulsom\index{Fulsom, B.}\\
Pacific Northwest National Laboratory\\
902 Battelle Boulevard\\
Richland, WA 99352, USA.}
\bigskip\bigskip
\end{raggedright}

Talk presented at the APS Division of Particles and Fields Meeting (DPF 2017), July 31-August 4, 2017, Fermilab. C170731

\section{Abstract}
Quarkonium is the bound state of a heavy quark and its anti-quark counterpart. The study of this system has experienced a renaissance thanks to results from e+e- collider experiments, including discoveries of long-predicted conventional quarkonia, and unusual states consisting of four quarks. The Belle Experiment operated at KEK in Japan from 1999-2010. Analysis of the collected data continues to produce new findings. The Belle~II experiment is a substantial upgrade of both the Belle detector and the KEKB accelerator, aiming to collect 50 times more data beginning in 2018. This talk presented recent Belle results related to hadronic and radiative decays in the bottomonium system. It described the capabilities of Belle~II to explore these topics, with a particular focus on the physics reach of the first data, where unique opportunities exist to make an immediate impact in this area.

\section{Introduction}
Quarkonium is the bound state of a heavy quark and its anti-quark counterpart. Phenomenological models, nonrelativistic QCD, and lattice calculations predict a large spectrum of particles whose properties (\emph{e.g.} mass, width, spin, production and decay mechanisms, etc.) can be verified in experiment. There have been many recent discoveries in this field, including several first observations of long-predicted quarkonia, but also many apparent ``$XYZ$'' states that cannot be explained within the quarkonium framework \cite{qqbar_review}. This includes several particles with non-zero charge, which cannot be $q\overline{q}$. The nature of these new states is the source of much interest, and there are many hypotheses ranging from ``meson molecules'' (weakly bound states of two mesons, \emph{i.e.} $D^{0}\overline{D^{0*}}$), to ``tetraquarks'' (directly bound pairs of color-singlet diquarks), to perhaps simple kinematic effects representing a non-exotic nature.

\section{Results from the Belle Experiment}
The Belle Experiment operated at the asymmetric KEKB $e^{+}e^{-}$ collider in Tsukuba, Japan from 1999 to 2010, collecting approximately 1.1 fb$^{-1}$ of data at various center-of-mass energies to produce various bottomonium resonances (see Table \ref{tab:data}). The record instantaneous luminosity achieved was $2.1\times10^{34}$ cm$^{-2}$ s$^{-1}$. The Belle Collaboration consists of approximately 400 members and has produced close to 500 publications at the time of writing.

\begin{table}[b]
  \begin{center}
    \begin{tabular}{|l|c|c|}
      \hline Resonance & Luminosity (fb$^{-1}$) & Yield (M) \\
      \hline $\Upsilon(5S)$ & 121 & 36\\
      \hline $\Upsilon(4S)$ & 711 & 772\\
      \hline $\Upsilon(3S)$ & 3 & 12\\
      \hline $\Upsilon(2S)$ & 25 & 158\\
      \hline $\Upsilon(1S)$ & 6 & 102\\
      \hline Other & 100 & N/A\\
      \hline
    \end{tabular}
    \caption{Summary of data collected by the Belle Experiment.}
    \label{tab:data}
  \end{center}
\end{table}

\subsection{Search for XYZ states in $\Upsilon(1S)$ decays}
In experiment, the $XYZ$ states have been found to decay preferentially to final states involving $\pi$ and $q\overline{q}$. The $\Upsilon(1S)$ has a non-negligible branching fraction to inclusive $J/\psi$ and $\psi(2S)$ final states, inspiring a search for potential decays to $XYZ$ and $\pi$ using the $\Upsilon(1S)$ Belle data set. Many such final states were considered, including decays to $\pi^{+}\pi^{-}J/\psi$, $\pi^{+}\pi^{-}\psi(2S)$, $K^{+}K^{-}J/\psi$, $\phi J/\psi$, $\pi^{\pm}J/\psi$, $\pi^{\pm}\psi(2S)$, and $K^{\pm}J/\psi$. No evidence for decays to $XYZ$ states from $\Upsilon(1S)$ were found, and limits were set for 14 potential $XYZ$ decay modes. This work was recently published in \cite{xyz_1s}.

\subsection{Search for $0^{--}$ glueball in $\Upsilon(1S)$ and $\Upsilon(2S)$ decay}
Similar to the $XYZ$ states, a search was conducted for $\Upsilon$ decays to a gluonic bound state ``glueball'' candidate $G_{0}$. Using the Belle $\Upsilon(1S)$ and $\Upsilon(2S)$ data sets, $\Upsilon(1S,2S)$ decays to $\chi_{c1}G_{0}$ and $f_{0}(1285)G_{0}$ were considered. The analysis also used a $\chi_{b1}$ sample identified via the radiative decay $\Upsilon(2S)\to\gamma\chi_{b1}$ to search for decays of $\chi_{b1}$ to $J/\psi G_{0}$ and $\omega G_{0}$ final states. In all cases, no evidence for these decays was found, and limits were established for glueball candidates with masses of 2.80, 3.81, and 4.33 GeV/$c^{2}$ for a range of natural widths from 0 to 500 MeV/${c}$. The complete results can be found in \cite{g_1s2s}.

\subsection{Study of $\eta$ and $\pi\pi$ transitions in $\Upsilon(4S)$}
As noted previously, the $XYZ$ states have been associated with di-pion transitions in quarkonium decay. In the conventional quarkonium picture, hadronic $b\overline{b}$ transitions via $\eta$ are predicted to be suppressed. However, previous experiment results have indicated enhanced branching fractions for $\Upsilon(4S)\to\eta\Upsilon(1S)$ \cite{bbr_eta} and $\Upsilon(4S)\to\eta h_{b}(1P)$ \cite{bel_eta}. Belle analyzed $\Upsilon(4S)\to\pi^{+}\pi^{-}\Upsilon(1S,2S)(\mu^{+}\mu^{-})$ and $\Upsilon(4S,1D)\to\eta(\pi^{+}\pi^{-}\pi^{0})\Upsilon(1S)$ decays in 496 fb$^{-1}$ of $\Upsilon(4S)$ data. Results consistent with past experimental values were obtained, but an improved examination of the $\pi^{+}\pi^{-}$ invariant mass spectrum revealed the first indications of an intermediate $f_{0}(980)$ contribution to this transition. The $\eta$ transition measurement confirmed an enhancement compared to the $\pi\pi$ decay rate, seen previously in other experiments and in violation of theoretic predictions, but found no evidence for $\Upsilon(1D)\to\eta\Upsilon(1S)$ decays \cite{eta_pipi}.

\section{Belle II}
The Belle~II experiment is the next generation $B$-Factory, expanding upon the successes of Belle. The goal of the experiment is to achieve approximately 50 times as much data as Belle, in order to search for new physics beyond the Standard Model. The collaboration currently consists of more than 750 members at 101 institutions representing 23 nations. Within the US, there are approximately 80 members at 14 locations.

In order to reach these goals, the KEKB accelerator is being upgraded to SuperKEKB, leading to a factor of 40 increase in instantaneous luminosity. This is accomplished largely by improved final focusing magnets capable of creating a compressed ``nano-beam'' interaction point, and also by a doubling in the beam current. The collisions are planned to occur mainly at the $\Upsilon(4S)$ center-of-mass energy, with a 7 GeV electron and 4 GeV positron beam.

To cope with the higher rates and backgrounds, the Belle detector is being updated with many new components. Vertexing is being replaced by a two-layer DEPFET and four-layer silicon pixel and strip detector. Charged-particle tracking will be mainly done with an improved small-cell drift chamber. The particle identification system has been replaced with a completely new and novel device in the barrel region, the ``imaging Time Of Propagation'' (iTOP) detector, and with a focusing aerogel ring-imagining Cherenkov detector in the forward endcap. The iTOP will use precise timing of the transmission of internally-reflected Cherenkov light from a quartz-bar radiator primarly to distinguish $\pi$ and $K$ tracks. Calorimetry will be provided by the same Belle CsI(Tl) electromagnetic calorimeter with upgraded electronics. Lastly, inner and foward endcap layers of the muon and $K_{L}$ detector in the flux return of the 1.5-T magnet have been upgraded with plastic scintillator to replace existing resistive plate chambers.

An accelerator commissioning phase (``Phase 1'') was completed with first turns of the SuperKEKB beam in 2016. The Belle~II detector was moved into position on the beam axis in April 2017. The next milestone is another period of commissioning from December 2017 through July 2018. This ``Phase 2'' will occur without the final vertex detector, and will feature the first $e^{+}e^{-}$ collisions in order to study background conditions and potentially produce some physics results with limited ($10\pm10$ fb$^{-1}$) data. The nominal Belle~II start with the complete detector (``Phase 3'') is scheduled to begin in the autumn of 2018. The ultimate target of the experiment is to collect 50 ab$^{-1}$ of data by the middle of the next decade.

With existing $B$-Factory data sets of ${\sim}1.5$ ab$^{-1}$, Belle~II is considering the opportunity for other physics results in Phase 2 and 3 before this integrated luminosity is reached. There is much that can be done in the context of quarkonium physics. Data collected below $\Upsilon(4S)$ opens the possibility to search for missing and study existing bottomonium states, and to look for new physics in bottomonium decay. With an energy scan of the correct region, direct production of the $D$-wave $\Upsilon$ states may be possible. Above the $\Upsilon(4S)$ there exists a unique opportunity to study exotic four-quark states, as less than 6 fb$^{-1}$ of data have been accumlated at Belle in this region.

\subsection{Physics Potential at $\Upsilon(6S)$}
The exotic charged $Z_{b}^{\pm}(10610,10650)$ states were discovered via $\Upsilon(5S)\to\pi\pi b\overline{b}$ transitions at Belle \cite{bel_zb}. A similar study at energies near 11 GeV showed evidence for $\Upsilon(6S)\to\pi\pi h_{b}(nP)$ (where $n=1$ and $n=2$), via $\pi Z_{b}^{\pm}(10610)$ decay \cite{bel_y6s}. Additional data at this energy would allow for studies of the nature of the $\Upsilon(6S)$ state, exotic quarkonia, and additional phase space to enable bottomonium discoveries. Of prime interest would be understanding $\Upsilon(6S)\to Z_{b}$ decay via charged and neutral dipion transitions to $h_{b}(1P,2P)$ and $\Upsilon(1S,2S,3S)$ final states. With indications that the $Z_{b}$ states are $B^{(*)}\overline{B}^{(*)}$ molecules, there are potential partner states (``$W_{b}$'') that could be discovered via decays such as $\Upsilon(6S)\to\gamma W_{b0}$ and $\rho W_{b0}$ with $W_{b0}\to \eta_{b}\pi$, $\chi_{b}\pi$, and $\rho\Upsilon$ \cite{th_voloshin}.

In analogy to the $\Upsilon(5S)$ evidence and observations at Belle, many new quarkonia transitions can be examined in order to further understand the nature of the above-threshold $\Upsilon$ states. These include $\Upsilon(6S)$ decays to $\pi^{+}\pi^{-}\Upsilon(n^{3}D_{J})$, $\eta\Upsilon(pS,n^{3}D_{J})$, $\omega\chi_{b}(1P)$, and $K^{+}K^{-}\Upsilon(1S)$. At ${\sim}11$ GeV, the $\Upsilon(6S)$ phase space could open the possibility for first discoveries of the $h_{b}(3P)$, $\Upsilon(2D)$, and $1F$ bottomonium multiplet via dipion or $\eta$ transition.

\subsection{Other points of interest above $\Upsilon(4S)$}
Collisions at center-of-mass energies near the other $B^{*}B^{*}$ thresholds also show potential for new discovery. Comparison of the hadron-to-$\mu$-pair cross section ($R$) versus energy with that for $\pi\pi\Upsilon$ transitions shows a pronounced dip \cite{bbr_scan} in the former where there is a significant rise in the latter \cite{bel_scan}. These features are similar to activity observed at the charm thresholds and near the $Y(4260)$ state. Perhaps this is an indication of an equivalent $Y_{b}$ state. Previous energy scans in this region by Belle and BaBar collected less than 1 fb$^{-1}$ of data, which could be surpassed in short order with the high-luminosity SuperKEKB.

\subsection{Below $\Upsilon(4S)$ bottomonium physics}
Below the $\Upsilon(4S)$ energy, both conventional $b\overline{b}$ and new physics can be studied. Of interest would be to study of the currently lesser-known $\Upsilon(1^{3}D_{J})$ triplet (for which two of the three states have yet to be observed) and $\eta_{b}(1S,2S)$ singlet states. Physics beyond the Standard Model could be manifested via enhancements to $\Upsilon(1S,2S)\to\chi\overline{\chi}$ ``invisible'' decays \cite{th_inv}, a $\chi_{b0}\to\tau^{+}\tau^{-}$ light Higgs coupling \cite{th_carl}, or dark sector decays of the type $\Upsilon(nS)\to\gamma\chi\overline{\chi}$ \cite{th_ds}. This region could be probed by collecting at least 200 fb${-1}$ of $\Upsilon(3S)$ data, approximately seven times the BaBar data set.

\section{Summary}
The field of quarkonium and ``new states'' is very active with many recent particle discoveries. It has led to many unexplained puzzles related to the nature of exotic tetraquarks and conventional $q\overline{q}$ states. The Belle experiment continues to provide relevant results in this area, with the measurement of $\Upsilon(4S)\to\pi^{+}\pi^{-}/\eta b\overline{b}$ transitions, and limits on searches for $XYZ$ production. SuperKEKB and the Belle~II experiment hold promise as the next generation $B$-Factory. Commissioning of the accelerator and detector will take place in the autumn of 2017, with a nominal experiment start date in 2018. The goal of Belle~II is to collect 50 times as much data as its predecessor, using an improved detector to search for new physics. It also offers a unique potential for further understanding of exotic hadrons and quarkonium.


\begin{thebibliography}{99}

\bibitem{qqbar_review}
Summarized in ``Developments in heavy quarkonium spectroscopy'' in K.~A.~Olive \emph{et al.} (Particle Data Group), Chin. Phys. C {\bf 38}, 090001 (2014).

\bibitem{xyz_1s}
C.~P.~Shen \emph{et al.} (Belle Collaboration), Phys. Rev. D {\bf 93}, 112013 (2016).

\bibitem{g_1s2s}
S.~Jia \emph{et al.} (Belle Collaboration), Phys. Rev. D {\bf 95}, 012001 (2017).

\bibitem{bbr_eta}
B.~Aubert \emph{et al.} (BaBar Collaboration), Phys. Rev. D {\bf 78}, 071103 (2008).

\bibitem{bel_eta}
U.~Tamponi \emph{et al.} (Belle Collaboration), Phys. Rev. Lett. {\bf 115}, 142001 (2015).

\bibitem{eta_pipi}
E.~Guido \emph{et al.} (Belle Collaboration), Phys. Rev. D {\bf 96}, 052005 (2017).

\bibitem{bel_zb}
A.~Bondar \emph{et al.} (Belle Collaboration), Phys. Rev. Lett. {\bf 108}, 122001 (2012).

\bibitem{bel_y6s}
R.~Mizuk \emph{et al.} (Belle Collaboration), Phys. Rev. Lett. {\bf 117}, 142001 (2016).

\bibitem{th_voloshin}
M.~B.~Voloshin \emph{et al.}, Phys. Rev. D {\bf 84}, 031502 (2011).

\bibitem{bbr_scan}
B.~Aubert \emph{et al.} (BaBar Collaboration), Phys. Rev. Lett. {\bf 102}, 0212001 (2009).

\bibitem{bel_scan}
D.~Santel \emph{et al.} (Belle Collaboration), Phys. Rev. D {\bf 93}, 011101 (2016).

\bibitem{th_inv}
G.~K.~Yeghiyan, Phys. Rev. D {\bf 80}, 115019 (2009).

\bibitem{th_carl}
S.~Godfrey and H.~E.~Logan, Phys. Rev. D {\bf 93}, 055014 (2016).

\bibitem{th_ds}
R.~Essig, P.~Schuster, and N.~Toro, Phys. Rev. D {\bf 80}, 015003 (2009).


\end{thebibliography}
\end{document}